\input psfig
\documentstyle[twoside,fleqn,espcrc2]{article}

\newcommand{\AmS}{{\protect\the\textfont2
  A\kern-.1667em\lower.5ex\hbox{M}\kern-.125emS}}

\title{Chiral Effects of Quenched $\eta'$ Loops
 \thanks{Talk presented by H.~Thacker}}
\author{ W.~Bardeen\address{ Fermilab, PO Box 500, Batavia, IL60510},%
 A.~Duncan\address{Dept. of Physics and Astronomy,University of Pittsburgh,
 Pittsburgh, PA 15260},%
  E.~Eichten$^{\rm a}$,%
  and 
  H.~Thacker\address{Dept. of Physics, University of Virginia, 
 Charlottesville, VA 22901}}
\begin{document}
\begin{abstract}
Preliminary results of a study of quenched chiral logarithms at $\beta=5.7$ are presented. Four 
independent determinations of the quenched chiral log parameter
$\delta$ are obtained. Two of these are from estimates of the
$\eta'$ mass, one from the residue of the hairpin diagram and
the other from the topological susceptibility combined with the Witten-Veneziano
formula. The other two determinations of $\delta$ are from measurement
of virtual $\eta'$ loop effects in $m_{\pi}^2$ vs. quark mass and in the chiral
behavior of the pseudoscalar decay constant. All of our results
are consistent with $\delta = .080(15)$. The expected absence of quenched
chiral logs in the axial-vector decay constant is also observed.
\end{abstract}

\maketitle

\section{Quenched Chiral Logs}

One of the most distinctive physical effects of light quark loops in
QCD is the screening of topological charge, which is responsible for the
large mass of the $\eta'$ meson. The absence of 
screening in the quenched approximation gives rise to singular
chiral behavior (quenched chiral logs) \cite{Sharpe} arising 
from soft $\eta'$ loops. In full QCD $\eta'$ loops are finite in
the chiral limit, while in quenched QCD they produce logarithmic
singularities which exponentiate to give anomalous power behavior
in the limit $m_{\pi}^2\rightarrow 0$. The first evidence of quenched 
chiral logs in the pion mass was reported last year\cite{lat98,CPPACS}. 
Here, and in a forthcoming paper \cite{future}, 
we present the results of a more detailed study of this effect.

In the context of the effective
chiral Lagrangian for QCD, the $U(3)\times U(3)$ chiral field $U=
\exp\left[i\sum_{i=0}^8\phi_i\lambda_i/f_{\pi}\right]$ exhibits
singular chiral behavior in the quenched theory arising from logarithmically
divergent $\eta'=\phi_0$ loops:
\begin{equation}
\label{eq:chfield}
U\rightarrow \exp\left[-\langle\phi_0^2\rangle/2f_{\pi}^2\right]\tilde{U}
= \left(\frac{\Lambda^2}{m_{\pi}^2}\right)^{\delta}\tilde{U}
\end{equation}
where $\tilde{U}$ is finite in the chiral limit, and $\Lambda$ is an upper
cutoff on the $\eta'$ loop integral. (We consider, for simplicity, the
case of three equal mass light quarks.)
The anomalous exponent
$\delta$ is determined by the $\eta'$ hairpin mass insertion $m_0^2$,
\begin{equation}
\label{eq:del}
\delta = \frac{m_0^2}{48\pi^2 f_{\pi}^2}
\end{equation}
(Here $f_{\pi}$ is normalized to a phenomenological value of 95 MeV.)
From the chiral behavior (\ref{eq:chfield}) of the field $U$, we can infer the 
singularities expected in the matrix elements of various quark bilinears.
Matrix elements of operators which flip chirality, such as the pseudoscalar charge
$\bar{\psi}\gamma^5\psi\propto U-U^{\dag}$ should exhibit a chirally singular
factor $(m_{\pi}^2)^{-\delta}$. By contrast, operators which preserve
chirality, such as the axial vector current $\bar{\psi}\gamma^5\gamma^{\mu}\psi
\propto iU^{-1}\partial^{\mu}U + h.c$ are expected to be finite in the chiral
limit. From PCAC, it follows that
\begin{equation}
\label{eq:chpi}
m_{\pi}^2 \propto m_q^{\frac{1}{1+\delta}}
\end{equation}
In the data presented here, we verify all of these predictions
by studying the pion mass and the vacuum-to-one-particle matrix elements
\begin{eqnarray}
\label{eq:fP}
\langle 0|\bar{\psi}\gamma^5\psi|\pi(p)\rangle &=& f_P\\
\label{eq:fA}
\langle 0|\bar{\psi}\gamma^5\gamma^{\mu}\psi|\pi(p)\rangle &=& f_A p^{\mu}
\end{eqnarray}

\begin{figure}
\vspace*{3.6cm}
\includegraphics{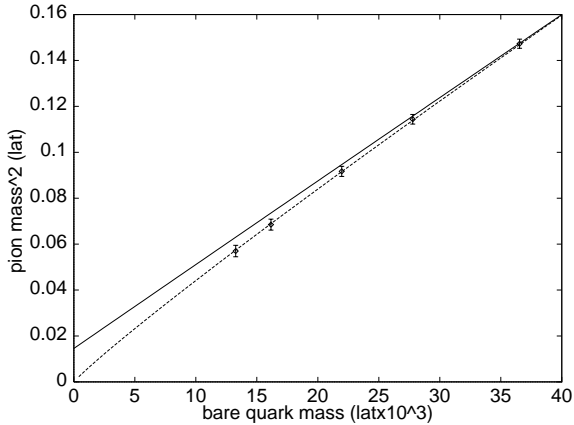}
\vspace{1.5cm}
\caption[]{Chiral log effect in $m_{\pi}^2$ vs. $m_q$. Solid line is a perturbative 
(quadratic) fit to the five largest masses (four are not shown). Dotted line is fit of all
masses to Eq. (3).}
\end{figure}

The modified quenched approximation (MQA) \cite{MQA} provides a practical
method for resolving the problem of exceptional configurations, and allows
an accurate investigation of the chiral behavior of quenched QCD with Wilson-Dirac
fermions. The pole-shifting prescription for constructing improved quark
propagators is designed to remove the displacement of real poles in the 
quark propagator, which is a lattice artifact, while retaining the contribution
of these poles to continuum physics. We have 
used the MQA procedure to study the masses and matrix elements of flavor
singlet and octet pseudoscalar mesons in the chiral limit of quenched QCD. 
A complete discussion of these results will be presented in a forthcoming
publication. Here we present results from 300 gauge configurations at $\beta=5.7$
on a $12^3\times 24$ lattice, with clover-improved
quarks ($C_{sw}=1.57$) at nine quark mass values covering a range of pion masses from
.2387(53) to .5998(17).

In addition to observing $\eta'$ loop effects, we also present two 
direct estimates of the chiral log parameter $\delta$, one from a calculation
of the $\eta'$ hairpin diagram, and the other from a calculation of the
topological susceptibility combined with the Witten-Veneziano relation.
The results for $\delta$ agree well with each other and with the exponents
extracted from $f_P/f_A$ and $m_{\pi}^2$, thus giving four independent and
consistent determinations of $\delta$. All four results fall within one standard
deviation of $\delta=.080(15)$.
This is about a factor of two smaller than the result $\delta=.17$ expected from the
phenomenological values of $m_0\approx 850$ MeV and $f_{\pi}=95$ MeV.
The agreement between the four
determinations of $\delta$ indicates that relations imposed by chiral symmetry and the
Witten-Veneziano formula are approximately valid, even at $\beta=5.7$. Possible reasons
for the overall suppression of the exponent $\delta$ compared to phenomenological
expectations will be discussed in Ref. \cite{future}.

\section{The $\eta'$ hairpin mass insertion}

Following Ref. \cite{Kuramashi} we calculate $\gamma^5$ quark loops using quark propagators
with a source given by unit color-spin vectors on all sites. 
The statistical errors are dramatically improved
by the MQA procedure, allowing a detailed study of the time-dependence even at the
lightest masses.
We determine the value of $m_0^2$ with a one-parameter fit to the overall magnitude of the hairpin correlator,
assuming a pure double-Goldstone pole form.
In the chiral limit, we find $m_0a=.601(30)$, or $m_0=709(35)$ MeV using
$a^{-1}=1.18$ GeV. (For unimproved $C_{sw}=0$ fermions, we get a much smaller
value of 464(24) MeV.)
Using Eq. (\ref{eq:del}), we 
obtain the values for the chiral log parameter $\delta$. 
Extrapolating to  the chiral limit, this gives
$\delta= .068(8)$ if we use the unrenormalized lattice value for $f_{\pi}$. If a
tadpole improved renormalization factor is included, this becomes $\delta=.095(8)$.

\section{Topological susceptibility}

The fermionic method for calculating the topological susceptibility\cite{Smit}
can be implemented with the same allsource propagators used for the hairpin
calculation. For each configuration, we compute the integrated pseudoscalar
charge $Q_5$. The winding number $\nu=-im_qQ_5$ is then obtained and the
ensemble average $\langle \nu^2\rangle$ is calculated. In the chiral limit,
this gives $\chi_t=(188\;MeV)^4$. (Here we have used $a^{-1}=1.18$ GeV.) 
Using the Witten-Veneziano formula (with unrenormalized $f_{\pi}$) to 
obtain $m_0$, the chiral log parameter $\delta = .065(8)$ is found.

\section{Quenched chiral logs in the pion mass}

The pion masses are obtained for nine kappa values over a range of hopping
parameters from .1400 to .1428, with $C_{sw}=1.57$, corresponding to quark masses
from roughly the strange quark mass down to about four times the up and down 
quark average (i.e. a pion mass of $\approx 270$ MeV). A perturbative 
(linear+quadratic) fit of
$m_{\pi}^2$ as a function of $m_q$ works well for the five heaviest masses,
up to $\kappa=.1420$. The value of $m_{\pi}^2$ for the lighter quark masses
deviates significantly below this perturbative fit (see Fig. 1) showing clear evidence
of a quenched chiral log effect. Fitting to the formula (\ref{eq:chpi}) gives
an anomalous exponent $\delta=.079(8)$.

\begin{figure}
\vspace*{3.6cm}
\includegraphics{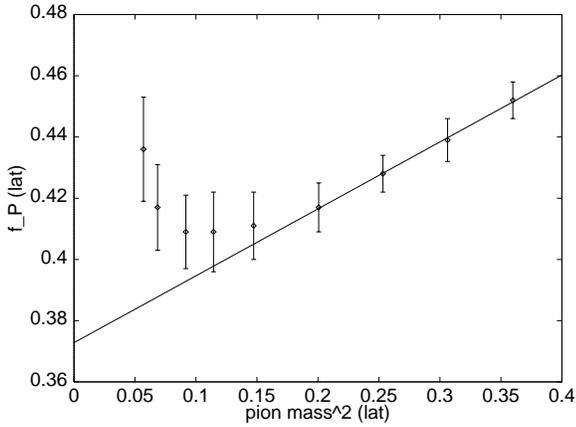}
\vspace{1.5cm}
\caption[]{$f_P$ vs. $m_{\pi}^2$ in lattice units. Solid line is a linear fit 
to the four largest masses.}
\end{figure}

\begin{figure}
\vspace*{3.6cm}
\includegraphics{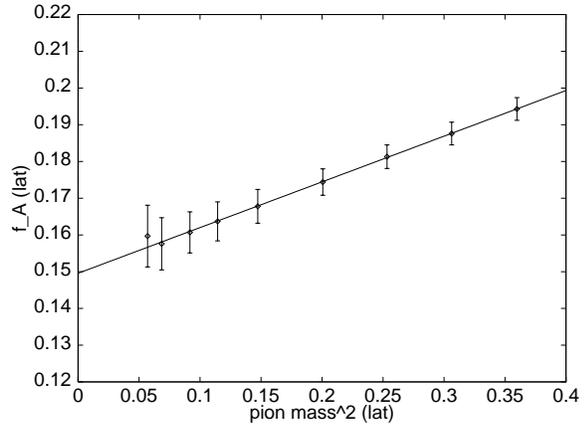}
\vspace{1.5cm}
\caption[]{$f_A$ vs. $m_{\pi}^2$ in lattice units. Solid line is a linear fit           
to the four largest masses.}
\end{figure}
 
\section{Pseudoscalar and axial-vector matrix elements}

By a combined fit of smeared-local pseudoscalar and axial-vector propagators,
we obtain values for the decay constants $f_P$ and $f_A$ defined in 
(\ref{eq:fP})-(\ref{eq:fA}). The behavior of these two constants as a function of pion
mass squared is shown in Figs. 2 and 3. We see that there is a very significant
chiral log enhancement at light masses for the pseudoscalar constant, but
the axial-vector shows no chiral log effect, just as theoretical arguments
predicted. From the phenomenology of $O(p^4)$ terms in the chiral Lagrangian,
it can be argued that the {\it perturbative} slopes of $f_P$ and $f_A$ in full
QCD should be approximately equal. Indeed, if a chiral log factor is removed from $f_P$,
the remaining slopes seen in our lattice data are equal within errors.
To obtain an estimate of $\delta$, we fit the ratio $f_P/f_A$ to a pure 
chiral log factor:
\begin{equation}
\frac{f_P}{f_A} = {\rm const.}\times (m_{\pi}^2)^{-\delta}
\end{equation}
This fit gives $\delta=.080(7)$.

\vspace{-0.1in}


\end{document}